\documentclass[aip, apl, 10pt, twocolumn,unsortedaddress]{revtex4-1}
\usepackage{graphicx}

\begin{document}

\title{Rotational tuning of interaction in metamaterials}

\author{Kirsty Hannam}
\author{David A. Powell}
\email{david.a.powell@anu.edu.au}
\author{Ilya V. Shadrivov}
\author{Yuri S. Kivshar}

\affiliation{Nonlinear Physics Centre, Research School of Physics and Engineering, Australian National University, Canberra, ACT 0200, Australia}

\begin{abstract}
We experimentally observe the tuning of metamaterials through the relative rotation of the elements about their common axis.
In contrast to previous results we observe a crossing of resonances, where the symmetric and anti-symmetric modes
become degenerate. We associate this effect with an interplay between the magnetic and electric near-field interactions and verify this by calculations based on the interaction energy between resonators.
\end{abstract}

\maketitle

Metamaterials created as an array of sub-wavelength, resonant elements can exhibit interesting electromagnetic properties, such as a negative refractive index~\cite{Smithetal2000}.  Unlike the atoms in natural materials, the near-field patterns of metamaterial elements are quite complex, giving rise to strong interactions between them.  Understanding this interaction is essential as it determines the overall resonant properties
and effective parameters of the material.  By controlling the relative arrangement of elements, it is possible to change this coupling and tune the properties of the structure~\cite{Liuetal2009,Lapineetal2009,Powelletal2010}. This allows us to alter the response of the material substantially without having to greatly alter the geometry or constituents.

\begin{figure}[b]
	\centering
		\includegraphics[width=\columnwidth]{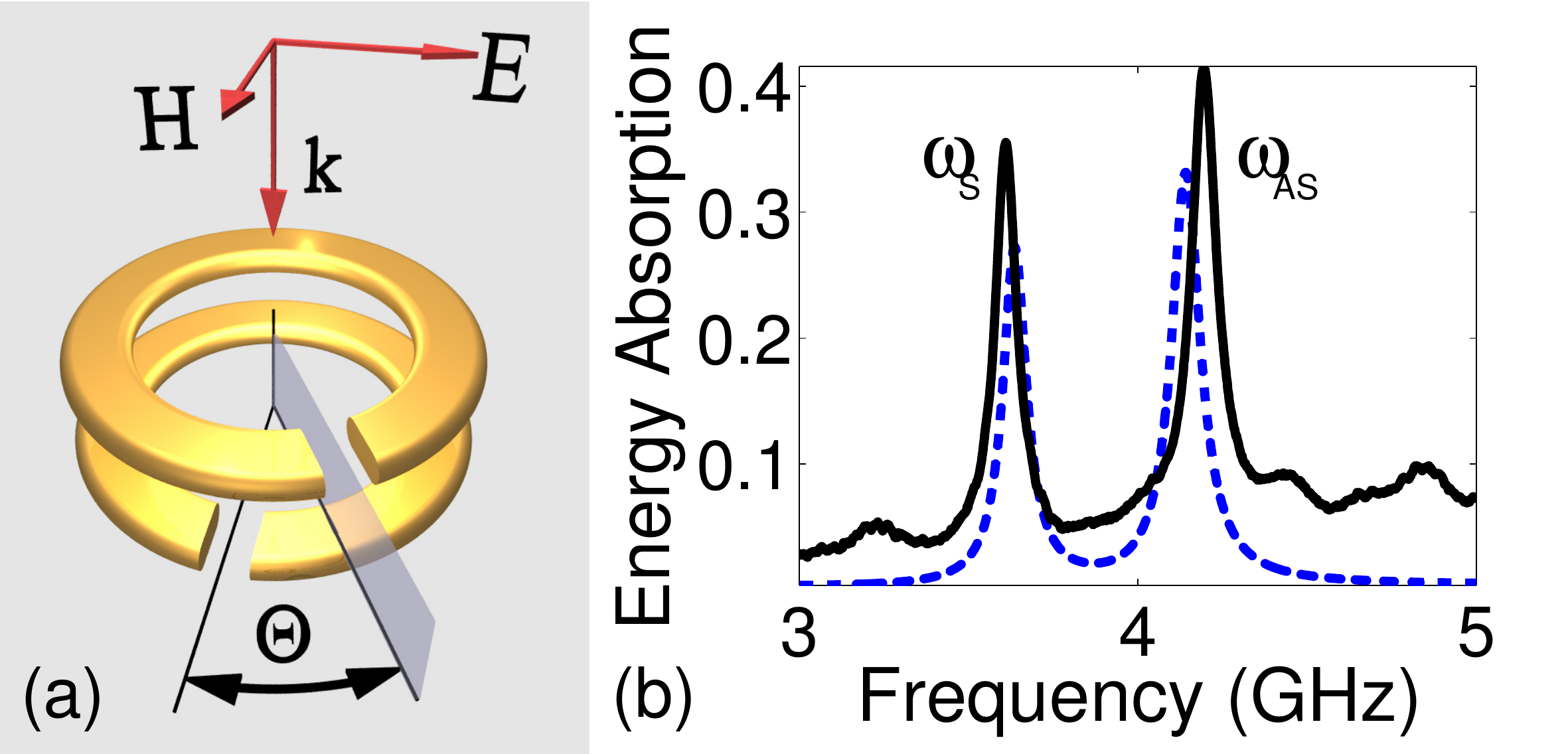}
	\caption{(a) A schematic showing the rings rotated with respect to each other through angle $\theta$, and the polarization of the incoming waves.  (b) A comparison of the experimental (solid) and numerical (dashed) absorption for angle $\theta = 90^\circ$.}
	\label{fig:Schematic}
\end{figure}

An important {\em building block} of metamaterials is the Split Ring Resonator (SRR)~\cite{Pendryetal1999}, which can exhibit a negative magnetic response due to its strongly resonant magnetic polarizability.  By coupling a pair of SRRs, chiral properties \cite{Liuetal2007} and trapped dark modes\cite{Li2009a} can also be observed.  In this Letter, we study the dynamics of two microwave SRRs broadside coupled to each other, varying twist angle $\theta$ between them, as shown in Fig.~\ref{fig:Schematic}(a).  A similar system operating at near infra-red frequencies was presented in Ref.~\onlinecite{Liuetal2009}, however in this work we show a substantially different regime of interaction verified experimentally for microwaves. 

The rings used have an inner radius of 3.5mm, an outer radius of 4mm, and a gap of 1mm.  They are copper, printed onto 1.6mm thick FR4 circuit board, and the rings are 3.6mm apart, with the dielectric boards located between the rings.  The incoming microwaves are polarized so that the electric field is in the $x$ direction, as shown in Fig.~\ref{fig:Schematic}(a).  Numerical calculations were performed using CST Microwave Studio with the boards having a dielectric constant of 4.6.

The most meaningful definition of the resonant frequency is the frequency of maximum excitation of currents within the rings.  This can be determined most readily by considering the absorption of the system, given by $1 - |S_{21}|^{2} - |S_{11}|^{2}$,
where $S_{21}$ is the transmission coefficient, and $S_{11}$ is the reflection coefficient.  A comparison of the experimental and numerical curves for $\theta = 90^\circ$ is shown in Fig.~\ref{fig:Schematic}(b).

\begin{figure}[b]
	\centering
		\includegraphics[width=\columnwidth]{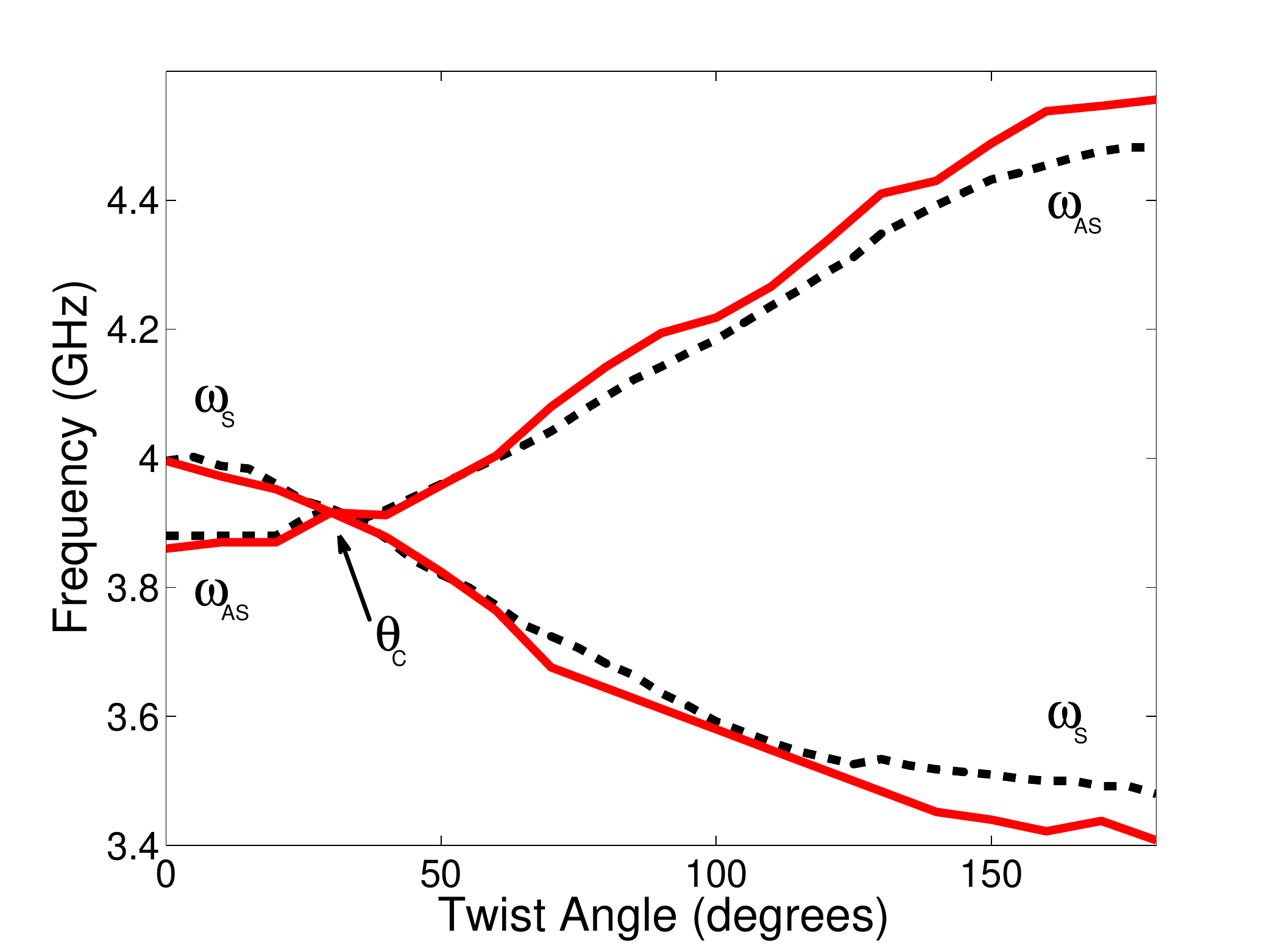}
	\caption{A comparison of the experimental (solid line) and numerical (dashed line) resonant frequencies.}
	\label{fig:Twist}
\end{figure}

Experimental results were measured using a Rohde and Schwarz ZVB network analyzer in a WR-229 rectangular waveguide, with $\theta$ varied from $0^\circ$ to $180^\circ$ in $10^\circ$ increments, while numerical results were calculated in $5^\circ$ increments.  The resonant frequency for each angle was found from the maximum of the absorption curve, and the resulting numerical and experimental results are compared in Fig.~\ref{fig:Twist}.

For $\theta = 0^\circ$, there are two resonances $\omega_{S}$ and $\omega_{AS}$, and by inspection of the currents in the rings we verify that these correspond to the expected symmetric and anti-symmetric modes.  As $\theta$ increases, $\omega_{AS}$  increases and $\omega_{S}$ decreases, reaching their maximum and minimum values respectively at $\theta=180^{\circ}$.  For our chosen parameters the resonances cross at $\theta_{c}\approx 33^{\circ}$, in contrast to Ref.~\onlinecite{Liuetal2009}, where an avoided crossing of resonances was found which was attributed to the electric quadrupole and octupole moments of the rings.

\begin{figure}[bt]
	\centering
		\includegraphics[width=\columnwidth]{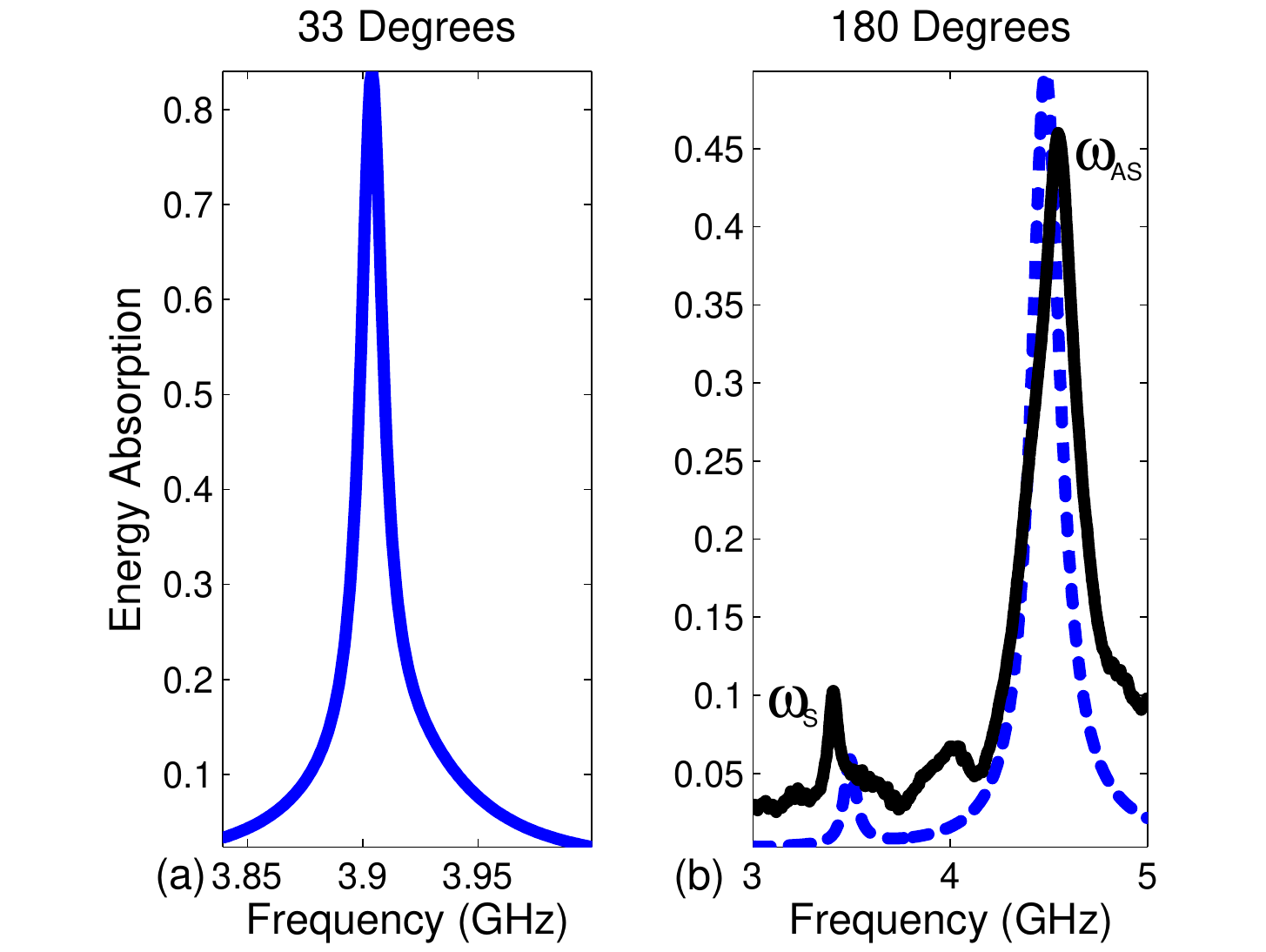}
	\caption{(a) The numerical absorption curve for $\theta_c$ ($33^\circ$), and (b) the numerical (dashed) and experimental (solid) absorption curves for $\theta = 180^\circ$.}
	\label{fig:crossing}
\end{figure}

The numerical absorption curve at $33^\circ$ is shown in Fig.~\ref{fig:crossing}(a).  This curve was calculated assuming loss free dielectric boards, so as to ensure the position of the crossing, given the low coupling of the asymmetric mode close to $\theta_c$.  It is quite clear here that there is only one resonance to be observed.

As can be seen in Fig.~\ref{fig:crossing}(b), the two resonances have different Q factors (widths), which can be attributed to conduction and radiation losses.  The suppression of radiation losses in the symmetric mode is related to the magnetic and electric dipole arrangements\cite{Li2009a}, where the electric dipole has much stronger radiation losses.  The asymmetric mode has mostly an electric dipole response, causing greater radiation losses, whereas the radiation losses at the symmetric mode have been suppressed by a dominating magnetic dipole.  The radiation distribution then changes with angle $\theta$, as can be seen by comparing Figs.~\ref{fig:Schematic}(b) and~\ref{fig:crossing}(b).
%

The tuning of the system by rotation can be explained by looking at the interaction between the rings.  As the rings are twisted, the magnetic and electric near-fields of the two rings change, changing the coupling between the resonances.  This is approached theoretically using the Lagrangian for a pair of coupled SRRs~\cite{Powelletal2010}:
\begin{equation} \label{eq:Lagrangian}
\mathcal{L}=A(\dot{Q}_1^2+\dot{Q}_2^2+2\alpha
\dot{Q}_1\dot{Q}_2)-B(Q_1^2+Q_2^2+2\beta Q_1Q_2)
\end{equation} 
where $\alpha$ and $\beta$ are the dimensionless magnetic and electric interaction constants and $Q(t)$ is the time-dependent mode amplitude.  By substituting Eq.~(\ref{eq:Lagrangian}) into the Euler-Lagrange equation we find that
\begin{eqnarray}\label{eq:Q1Q2}
\ddot{Q}_1+\omega_0^2 Q_1=-\alpha \ddot{Q}_2-\beta \omega_0^2 Q_2.\\
\ddot{Q}_2+\omega_0^2 Q_2=-\alpha \ddot{Q}_1-\beta \omega_0^2 Q_1.
\end{eqnarray}
This then allows the two resonances to be found - symmetric (when $Q_1 = Q_2$), and asymmetric (when $Q_1 = -Q_2$):
\begin{equation}
\omega_{S} = \omega_{0}\sqrt{\frac{1 + \beta}{1 + \alpha}}, \qquad \omega_{AS} = \omega_{0}\sqrt{\frac{1 - \beta}{1 - \alpha}}.
\label{eq:omega_s_as}
\end{equation}

In principle if $\omega_0$ is known, then by inverting Eq.~(\ref{eq:omega_s_as}) it is possible to fit $\alpha$ and $\beta$ from $\omega_{S}$ and $\omega_{AS}$. However we found that this procedure is extremely sensitive to error and does not yield usable results.  Instead, we start from the approach outlined in Ref.~\onlinecite{Powelletal2010} to evaluate the interaction energy between the fundamental modes of the rings.

\begin{figure}[bth]
	\centering
		\includegraphics[width=\columnwidth]{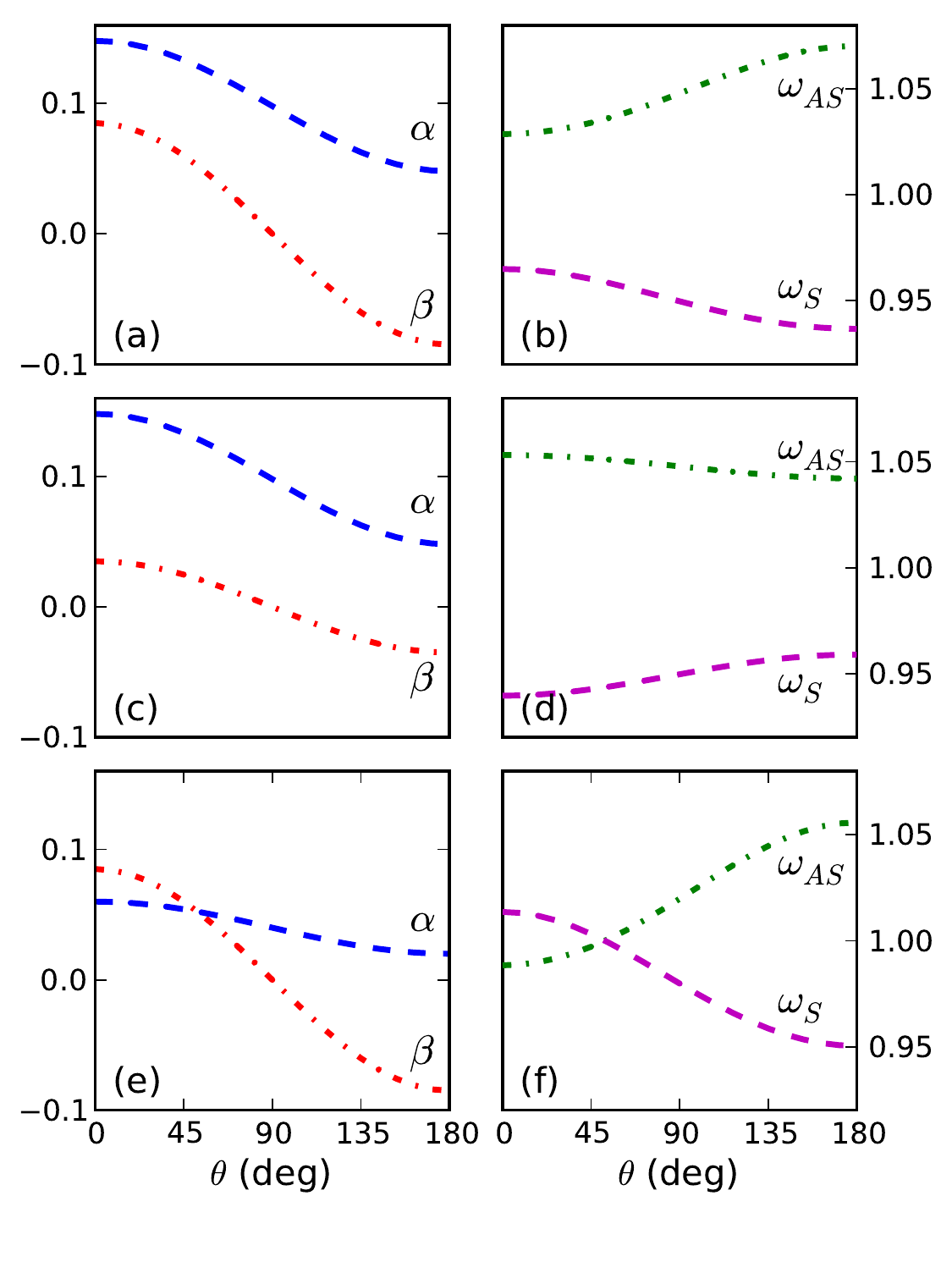}
	\caption{(a) Magnetic ($\alpha$) and electric ($\beta$) interaction constants calculated for a pair of rings in free space, and (b) corresponding resonant frequencies. (c) Interaction constants for case where electric coupling dominates at $\theta=0^\circ$, and (d) corresponding resonant frequencies.  (e) Interaction constants which become equal for some angle $\theta$, and (f) corresponding resonant frequencies.}
	\label{fig:interaction_all}
\end{figure}

For a pair of rings in {\em a homogeneous dielectric background}, the interaction constants are shown in Fig.~\ref{fig:interaction_all}(a). We find that the interaction constants are very well described by $\beta = \beta_{1}\cos(\theta)$ and $\alpha = \alpha_{0} + \alpha_{1}\cos(\theta)$ with $\beta_1 =0.085$, $\alpha_0 =0.098$ and $\alpha_1 =0.05$. These constants are dictated by the charge separation across the gap of the ring, the current circulating around the ring, and the inhomogeneity of the current distribution around the ring, respectively.  For rings aligned on the same axis, we expect that the magnetic interaction should always be positive, as the intersecting magnetic field from one loop should always be normal to the other loop.  In addition the electric interaction should be positive at $\theta=0^\circ$ as the charge distribution has the nature of parallel dipoles. All arrangement of rings on the same axis which we considered obeyed these considerations.

In Fig.~\ref{fig:interaction_all}(b) we plot the corresponding frequencies of the symmetric and anti-symmetric modes, normalized to $\omega_0$.  As our approach models the response of the resonators in a homogeneous dielectric background, the results are significantly different from those observed experimentally, where the dielectric is inhomogeneous and the effect of waveguide boundaries is also significant. In particular, the crossing of resonances cannot be reproduced for rings in homogeneous background. Therefore we consider the possible regimes of interaction which may occur, under the assumption that the interaction constants can be fitted as described above.

The case considered in Fig.~\ref{fig:interaction_all}(a-b) corresponds to the magnetic interaction always being larger than the electric interaction.  This results in increasing splitting of $\omega_S$ and $\omega_{AS}$ with increasing twist angle, however in principle there is no reason why the splitting cannot decrease.  We show such a case in Fig.~\ref{fig:interaction_all}(c-d), where we have set $\beta_{1}=0.02<\alpha_1$, such that the inhomogeneity in the current has a stronger influence than the dipole-like charge distribution.  Despite the apparent difference in frequency splitting curves, there is little difference between the interaction constants shown in Fig.~\ref{fig:interaction_all}(a) and (c).

The only other case allowed in our model of interaction under the afore-mentioned physical constraints on $\alpha$ and $\beta$ is that $\alpha > \beta$ for $\theta=0^\circ$.  An example of this is given in Fig.~\ref{fig:interaction_all}(e), where we have reduced the magnetic coupling such that at some angle $\alpha=\beta$, by setting $\alpha_0=0.04, \alpha_1=0.02$.  The corresponding resonant frequencies normalized to $\omega_0$ are plotted in Fig.~\ref{fig:interaction_all}(f).  We see that for low twist angle, $\omega_{S}$ occurs at a higher frequency, but decreases with angle and crosses $\omega_{AS}$.  Clearly this regime corresponds to what we observe in experiment, and we hypothesize that the inhomogeneous dielectric serves to enhance the electric interaction relative to the magnetic interaction in our system.

In Ref.~\onlinecite{Liuetal2009} it was claimed that the resonances converge, undergo an avoided crossing, then diverge, as $\theta$ increases. Utilizing both our analytical model and full numerical simulation, we could not find this regime in our system.  We also note that for our structure the assumption of constant magnetic interaction with twist angle is not justified.  However by taking the inhomogeneity of magnetic interaction into account, we find that we can neglect higher order electric interactions.

Equations~(\ref{eq:omega_s_as}) show that the tuning curves arise from competition between electric and magnetic interaction constants.  This means that even when interaction is strong, the corresponding frequency splitting can still be weak. The experimentally-observed crossing angle $\theta_c=33^\circ$ represents a particularly interesting case, and it occurs when $\alpha=\beta$.  In the lossless case the symmetric and anti-symmetric resonances become degenerate, and although both electric and magnetic interaction coefficients are strong, they effectively cancel each other out.  However in the presence of losses, we need to consider the eigenfrequencies $\omega_{S}$ and $\omega_{AS}$ as complex values, with imaginary parts corresponding to the losses.  By examining Fig.~\ref{fig:crossing}(b) we see that the coupling of the modes to the waveguide is not identical, hence they have different radiation losses.  Thus it becomes very unlikely that in the complex plane $\omega_{S}=\omega_{AS}$ for any parameter value, and true degeneracy does not occur.  Therefore it is not possible to cancel out the interaction between rings, and our numerical work verified this.

In conclusion, we have shown that by changing the relative rotation between two rings, we can significantly change the coupling, which causes the resonances to change.  We have found that there is a crossing where the two resonances coexist, which corresponds to equal electric and magnetic coupling.

\begin{acknowledgments}
 We acknowledge funding from the Australian Research Council and useful discussions with Maxim Gorkunov and Andrey Miroshnichenko.

\end{acknowledgments}

\end{document}